\documentclass[preprint,12pt]{elsarticle}

\usepackage{amssymb}

\journal{International Journal of Human-Computer Studies}

\begin{document}

\begin{frontmatter}



\title{Judgement and the Feeling of Agency When Instructing a Computer to Act}


\author[inst1]{Johanna Didion}
\author[inst1]{Diego Garaialde}
\author[inst1]{David Coyle}

\affiliation[inst1]{organization={School of Computer Science, University College Dublin},
            addressline={Belfield}, 
            postcode={Dublin 4}, 
            country={Ireland}}

\begin{abstract}
In cognitive psychology, the sense of agency refers to the feeling of controlling an external event through one’s own actions. This paper asks how people experience agency when they give an instruction to a computer which then acts on their behalf. Research in human-human interaction has found that while people experience agency over the initial actions taken by another in response to their command, they may not experience agency over the ultimate outcome of those actions. We present two experiments. The first draws on prior studies of human-to-human interaction but replaces a human co-actor with a computer. The second study addresses potential limitations in the first, exploring the effect of more meaningful outcomes, and assessing both pre-reflective feelings and post-reflective judgements of agency. We find that participants experienced agency over the initial action taken by the computer in response to their command, but did not experience pre-reflective agency over the final outcome.
Interestingly, however, they did judge themselves to have caused the final outcome, when it was meaningful and their command specific. This difference between the pre-reflective feeling and post-reflective judgement of agency has important implications for design. If designers want people to feel in control and take responsibility for outcomes that result from instructions given to a computer, they cannot rely on pre-reflective intuition. Instead, we should design systems that encourage people to reflect on their actions and make deliberative judgments regarding responsibility.
\end{abstract}

\begin{highlights}
\item We examine people’s judgement and feeling of agency when giving instructions to a computer which acts on their behalf.
\item 
People did not report pre-reflective feelings of agency for outcomes resulting from instructions given to a computer.
\item 
People did judge themselves to have caused outcomes when their instructions were both meaningful and specific.
\item 
Systems should encourage people to reflect on their actions and make deliberative judgments regarding responsibility.
\end{highlights}


\end{frontmatter}


\section{Introduction} \label{sec:Introduction}
Humans take hundreds of actions every day. Often these actions are 
goal-directed, aimed at changing the current state of their environment.  
The actions can be small, like pressing a switch to turn on a light. Or more significant, like quitting a job. But they all have outcomes. 
Within cognitive psychology the term \textit{sense of agency} is used to describe the experience a person has of controlling their actions and, through this control, affecting the external world \cite{Haggard2009, chambon2014action}. The sense of agency is a crosscutting experience essential to many aspects of our day-to-day lives. 
It links to concepts such as control, ownership, responsibility and causality, and has a significant impact on how we perceive our world \cite{Coyle2012}. The RDoC \cite{insel2010research}, a recent framework that aims to structure research on mental health around fundamental domains of human functions, identifies agency as a core subconstruct in both our sensorimotor and social systems. 

Interestingly, people's sense of agency is remarkably malleable and often fragile. In the case of mental health, this malleability can have negative consequences. In severe cases people can feel disassociated from their own thoughts and actions \cite{mellor1970first}. In other instances, people can feel an exaggerated sense of agency. \citet{Coyle2012} cite the example of gamblers placing higher bets when they, rather than someone else, initiate a bet due to an increased sense of control over the event. Even though the actual odds remain the same. They argue that malleability in the human sense of agency offers both opportunities and pitfalls in human-computer interaction (HCI).

In recent years an increasing number of HCI studies have addressed the sense of agency \cite{Limerick2014, bergstrom2022sense, cornelio2022sense}. Some have investigated different input modalities, suggesting for example, that while skin-based input can increase the sense of agency \cite{Coyle2012, BergstromLehtovirta2018}, spoken input may reduce it \cite{Limerick2014}. There is also a significant body of work investigating agency in the context of embodiment and user representation in virtual and augmented reality, and in regard to human augmentation \cite{kasahara2019preemptive, tajima2022whose, abtahi2022beyond, seinfeld2021user, banakou2014body, kilteni2012sense}. As a whole, this prior work has tended to examine agency from an individual perspective. The emphasis is on individual sensorimotor control and execution of goal-oriented behaviour in situations involving only one actor. This paper takes a different approach, considering agency in situations that involve more than one actor.

Events involving multiple actors are inherently more complex than those involving only one. For example, it may be difficult to distinguish the actions of one person from those of another, particularly when attempting to attribute responsibility for an outcome that is caused by joint action. Previous research in cognitive psychology has addressed this issue in the context of human-human interaction. Work by \citet{Pfister2014} and \citet{Caspar2018} investigated people's sense of agency in situations where one person issues a command to another person who then acts on that command. This perspective has not been investigated in HCI contexts, but we argue it is particularly relevant. Interactions with a computer often involve a person issuing an instruction to which a computer responds, potentially setting off a chain of events/outcomes. \citet{Pfister2014} found that while people feel agentic over the initial action taken by another person in response to their command, they may not feel agentic over the resultant outcome caused by the second person's action. In this paper, we examine this phenomenon in an HCI context, where a person issues a command to a computer which then acts on their behalf. The aim is to extend our understanding of agency in human computer interaction by considering situations involving human-computer co-action.

The paper presents two studies. The first study builds on the \citet{Pfister2014} study on human-to-human interaction but replaces a human co-actor with a computer co-actor. In the second study, we adapt and extend our initial paradigm, introducing a game-based task to make the experiment more meaningful to participants. We also investigate the impact of specificity in the instructions given to a computer and consider the difference between instinctive (pre-reflective) feelings of agency and subjective (post-hoc) judgements of agency. 

The paper makes several contributions. It is amongst the first in HCI to consider the sense of agency in a joint action context, considering both instinctive (pre-reflective) and judgement-based (post-reflective) measures. We show that people do not experience an instinctive sense of agency over outcomes that result from commands they issue to a computer. This lack of pre-reflective agency is consistent with prior studies of human-human interaction. However, we also show that under certain conditions people do judge themselves to be responsible for outcomes resulting from commands they give to a computer. This difference between pre-reflective instinct and post-reflective judgement has important implications for design. Agency is closely linked to responsibility. If we want people to feel in control and take responsibility for outcomes that result from instructions they give to a computer, we cannot rely on pre-reflective instinct. Instead, we should design systems that encourage people to reflect on their actions and make a deliberative judgment regarding responsibility.

\section{Related Work} \label{sec:RelWork}
\subsection{The Sense of Agency} \label{relWork:SoA}
\citet{Synofzik2008} describe a distinction between a non-conceptual feeling of agency and a conceptual judgement of agency.
The feeling of agency is a pre-reflective experience, happening unconsciously when we feel in control of our own voluntary actions and their outcomes. \citet{chambon2014action} describe this feeling of agency as a background experience. We are not always aware of it, but when our deliberate actions cause an outcome, it is this feeling of agency that allows us to intuitively know “I did that” \cite{Coyle2012}. \citet{chambon2014action} note that we are often most aware of this feeling of agency when our actions do not unfold as expected or fail to produce intended results, as this diminishes our sense of agency.   
In contrast, the judgement of agency arises consciously and after the event when we evaluate our own actions, or those of another, and attribute  responsibility or causality for an outcome. 
Both levels of the sense of agency are related but seem to result from different underlying processes in the human brain \cite{Dewey2014}. Other methods are also employed when measuring explicit judgements and the implicit feelings of agency.

\subsubsection*{Judgement of Agency}
Measurement of the post-reflective judgement of agency typically involves a rating scale or questionnaire asking a person whether they experienced agency, felt in control, or caused an outcome. Usually, participants make their judgements using a Likert style scale, but some studies ask for a yes/no response. The key point is that these measures are explicit retrospective judgements, as the questions are asked and answered after all actions and outcomes are completed.

\subsubsection*{Feeling of Agency}
In cognitive psychology, the pre-reflective feeling of agency is measured implicitly. The most widely used implicit measurement is \textit{temporal} or \textit{intentional binding} \cite{moore2012intentional, Engbert2007, Haggard2002}. This phenomenon was first reported by \citet{Haggard2002} and repeated studies have shown it to be a robust implicit measure of the sense of agency \cite{moore2012intentional}.
\citet{Haggard2002} found that when people perform a voluntary action which causes a clear outcome, they tend to experience a temporal binding effect, where the action and outcome are experienced as closer together in time than is actually the case. They further showed that when actions are involuntary, or where people do not feel responsible for an outcome, the temporal effect is reversed. Here the interval between events is perceived as longer than was actually the case. Given that the temporal binding effect occurs when actions are intentional, Haggard coined the term \textit{Intentional Binding} to describe the phenomenon. 

The use of intentional binding to implicitly measure people's sense of agency in HCI was first described by \citet{Coyle2012}. They provide a detailed description, and experimental examples, of the two main methods to measure intentional binding: the Libet Clock method, and interval estimation. Both methods have been widely used in cognitive psychology research. Each has strengths and limitations. The Libet Clock method provides a detailed breakdown of where binding occurs within an action-outcome sequence and is highly robust. But it is time-consuming to administer and is not well suited to experiments that involve complex visual tasks.
In contrast, interval estimation provides a less detailed breakdown of when temporal binding occurs, because there is no distinction between action and outcome binding. But it is significantly less time-consuming to apply and is more suited to experimental designs involving complex tasks or multiple experimental conditions. For these reasons, the experiments in this paper use the interval estimation approach.  

\subsection{The sense of agency in HCI} \label{relWork:SoAHCI}
When referring to the sense of agency in the context of human-computer interaction, the term usually describes the sense of being in control of when interacting with technology. Similar to actions taken in real life, actions taken with technology produce specific outcomes. 
These outcomes may happen in our real environment, for example turning the lights on using a mobile application, or within the technology itself, such as using a controller to move a character in a computer game. Researchers have argued that the sense of agency plays an important role in human-computer interaction as it can impact the overall user experience and enjoyment \cite{Limerick2014}, as well as the feeling of responsibility for the outcomes produced \cite{Moretto2011}.

In HCI literature the words agency and control are often used interchangeably. For example, recent research has investigated how the design of YouTube influences people's sense of control, contrasting curated playlists with autoplay features \cite{lukoff2021design}, and examined the impact of personalised recommender systems \cite{dean2020recommendations}. Control and responsibility have also been investigated in areas such as automated driving \cite{heikoop2019human} and decision-making systems \cite{douer2021decision}. In contrast, HCI research that builds on the neurocognitive perspective on agency outlined above has tended to focus on specific interaction methods.
It has been shown that different input modalities can improve or decrease the sense of agency. For example, tapping the own skin instead of a button may enhance the implicit sense of agency \cite{Coyle2012, BergstromLehtovirta2018} and using voice commands can diminish it \cite{Limerick2015}. Research has also investigated the impact of feedback modalities on agency, including work on tactile and mid-air haptic feedback \cite{cornelio2017agency, cornelio2022sense}. There is a significant body of work investigating agency in the context of embodiment and user representation in virtual and augmented reality and in regard to human augmentation \cite{kasahara2019preemptive, tajima2022whose, abtahi2022beyond, seinfeld2021user, banakou2014body, kilteni2012sense, Nostadt2020embodiment}. Finally, HCI researchers have developed novel approaches to measuring intentional binding, e.g. making use of audio channels to reduce the visual complexity of Libet Clock measurements \cite{cornelio2018beyond}.

As noted in our introduction, this previous HCI research has tended to focus on the sensorimotor experience of agency and typically on interactions involving one person. \citet{cornelio2022sense} offer a detailed definition of agency in this context, stating that it must involve individual intentional actions and a motor target, noting that this definition excludes technologies such as recommender systems and many autonomous systems. This definition is very helpful for many situations, but does not account for situations involving more than one person or situations when the computer is explicitly seen as a co-actor. In general such situations have received less attention within HCI research. This is not to say no work has been done. \citet{Coyle2012} describe an experiment in which the level of computer assistance people receive for a computer was manipulated. They found that while a low level of assistance supports agency, high levels of computer assistance, even when it helped users to achieve their goals, resulted in a lower sense of personal agency \cite{Coyle2012}. Similar results are reported by \citet{Berberian2012} in the context of automated air traffic control systems.  

\subsection{Joint Action and the Sense of Agency} \label{relWork:SoASocialContexts}
In contrast to individual sensorimotor agency, agency is also important in settings with more than one actor. Here the sense of agency includes perception and interpretation of both one's own and others' actions \cite{insel2010research}. Agency in joint actions has been researched more intensively in the fields of Neuroscience and Psychology. Here we discuss three broad threads within this work: vicarious agency, joint agency, and command-agent agency.

\textit{Vicarious agency} describes situations where a person experiences agency over actions taken by another person. \citet{Nielsen1963} conducted an experiment where participants did a drawing task. A mirror was placed in such a way that the participant observed someone else's hand performing the task instead of their own. Participants were not aware that they were not actually seeing their own hand, causing them to perceive the external hand as their own, even in situations where the movements did not align with those of their real hand \cite{Nielsen1963}. In a more recent study by \citet{Wegner2004}, an experimenter stood behind a participant while the participant kept their arms at their sides. 
Cardboard separated the experimenter and participant, who wore a smock. The experimenter reached their arms forward through the sleeves of the smock, allowing them to act as the arms of the participant. Participants reported a sense of control over the experimenter's arms, particularly when they heard instructions that previewed upcoming hand movements \cite{Wegner2004}. 

\textit{Joint agency} describes the experience of agency when two or more actors act together to complete a task \cite{Loehr2022}.
Joint agency covers the feeling of "We did that together" compared to the individual feeling of "I did that" or "You did that" \cite{Loehr2022}. \citet{Obhi2011a} report a study on agency in joint actions where two participants had to press a button to initiate a tone. Participants were instructed to either act first, or act as soon as the other person acted. 
On this occasion, while the initiator reported an explicit judgement of agency, both participants showed intentional binding, suggesting that they experienced a feeling of agency.  \citet{Obhi2011a} argue that the participants formed "a new agentic identity (a 'we' identity)". This identity resulted in a sense of joint agency, which has also been evaluated by \citet{Bolt2016}. They found that people experience a higher sense of joint agency for actions where they jointly coordinate their actions. 
Building on their studies of joint agency in human co-action \citet{Obhi2011} also evaluated joint agency in a study with a computer as a co-actor. They did not find the same implicit sense of agency found in the human-human condition for the human-computer condition. However, participants reported an explicit judgement of agency, when they were told, that they caused the outcome.

\textit{Command-agent agency} is the term we use to describe agency in situations where one person instructs or commands another person to act. 
\citet{Pfister2014} describe an experiment in which two participants took on the role of \textit{leader} and \textit{follower}. 
The leader initiated a task by pressing a button which caused an outcome. The follower was then required to press a second button, as quickly as possible and this again caused an outcome. The results provided evidence that the leader experienced a sense of agency for the follower's actions - as measured by intentional binding. 
However, neither participant experienced a sense of agency for the final outcome - the outcome of the follower's action. Hence, neither the leader nor the follower appears to experience a sense of agency over the ultimate outcome. 
A more recent study looking at this \textit{commander-agent} relationship found that a person can experience agency over the ultimate outcome when acting themselves compared to when they are giving instructions to an agent to act on their behalf \cite{Caspar2018}.
These studies suggest that giving instructions to another human does reduce the feeling of agency and responsibility for outcomes when compared to acting for oneself.

\section{Overview and Research Questions}  \label{sec:Overview}
The primary aim of this paper is to explore and extend our understanding of joint action in the realm of human-computer interaction. Our work builds on neuroscience and psychology research, and most closely on work on the command-agent perspective outlined above. We seek to understand how people experience agency when a computer follows their instructions and acts on their behalf. We believe that this perspective is particularly interesting in an HCI context, as interaction with a computer often involves a person issuing a command (taking an initial action), to which a computer then responds, potentially setting off a chain of events/outcomes. In many cases this involves simple day-to-day interactions, e.g. asking Apple's Siri or Amazon Alexa to play a song or turn on the lights. But it can also involve situations with significant consequences, such as AI assisted air traffic control systems or robotic medical and even military systems. We also argue that command-agent situations are likely to become both more common and increasingly complex as use of AI in computing systems becomes more widespread.

The following research questions are addressed: 
\begin{itemize}
\item Do people experience a sense of agency when a computer responds to an instruction they give?
\item Do people experience a sense of agency over the final consequences of an instruction given to a computer?
\item Is there a difference in people's feeling and judgement of agency over the final consequence of an instruction given to a computer? 
\item Does the specificity of the instruction and a more meaningful outcome have an impact on people's sense of agency?
\end{itemize}
Our questions are addressed through two studies.

\section{Study 1} \label{sec:study1}

Our first experiment was built on previous work in human-human interaction by \citet{Pfister2014}. As this experiment took place when COVID-19 restrictions were in place it was not possible to being participants into a lab. We therefore conducted our study online. Our design closely followed the design of \citet{Pfister2014}, but replaced human-human co-action  with a computer co-action scenario. We also extended the \citet{Pfister2014} study by considering different degrees of specificity in the instruction participants gave to a computer. As such, this study is best described as a theoretical replication \cite{yin2009case} of the study by \citet{Pfister2014}.

In running our study online we were aware of potential concerns around this approach. Indeed the HCI community as a whole has faced well-documented challenges over recent years, with researchers increasingly required to move studies online due to COVID-19 restrictions on in-person experiments \cite{9463290}. However, empirical studies by \citet{Galang2021} have demonstrated that online experiments on the sense of agency are valid and produce results consistent with in-person experiments. Similar to our paper, they conducted an asynchronous study implemented with the jsPsych Javascript library \mbox{\cite{de2015jspsych}}, and using intentional binding as an implicit measure of the sense of agency. This gave us confidence in the validity of our online approach.

Based on previous studies, our hypotheses were as follows:
\begin{enumerate}
    \item Participants will experience a sense of agency over the action taken by the computer, irrespective of the specificity of their instruction.
    \item Participants will experience a greater sense of agency over the final outcome when the instruction they give is specific.
\end{enumerate}

This study was given ethical permission by the University College Dublin Human Research Ethics Committee.

\subsection{Procedure}
\begin{figure}
    \centering
    \includegraphics[width=\linewidth]{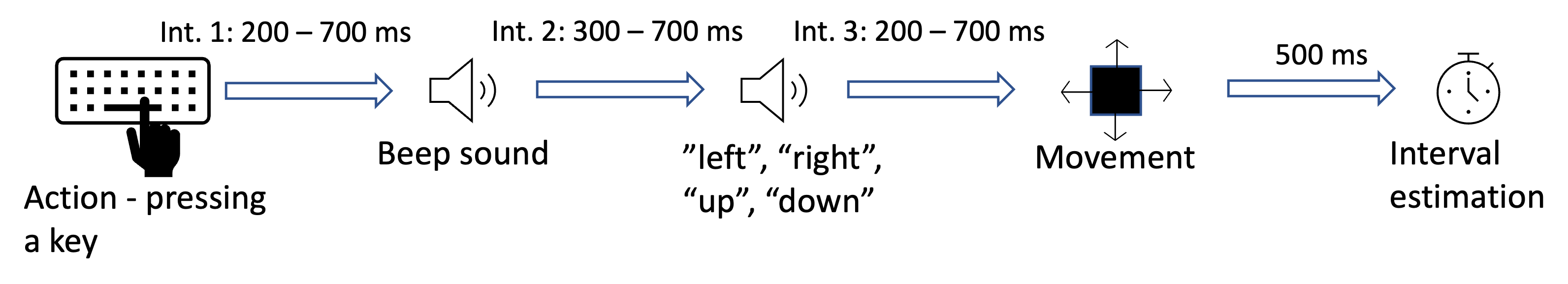}
    \caption{Procedure of each round - The participant starts by pressing a key on the keyboard, after 200 - 700 ms, this initiates a beep sound. After 300 - 700 ms, the computer reacts to this by giving a voice command indicating a direction (left, right, up or down). After another 200 - 700 ms, the object, that is seen on screen, moves in the indicated direction. 500 ms after the movement is completed, the participants see a screen, where they are asked to estimate the length of one of the three intervals.}
    \label{st1:fig:round}
\end{figure}

The overall procedure for each experimental trial is shown in Figure \ref{st1:fig:round}. It is based on the procedure of \citet{Pfister2014}, with several modifications\footnote{We must note in advance that we had some misgivings about the complexity of the experimental task used in this study. These misgivings are addressed further in the discussion section following this study and are also addressed by our second study. However, allowing for these misgivings, we decided that the most rigorous approach for our first experiment was to closely follow the procedures applied in previous research.}.

In the experiment, participants took an action by pressing a key on a keyboard. The experiment consisted of two conditions, which allowed to test the impact of the specificity of participants' actions. In condition 1 -  \textbf{\emph{the general condition}} - they pressed the spacebar. In condition 2 -  \textbf{\emph{the specific condition}} - they pressed one of the four arrow keys: up, down, left, or right. The human follower (co-actor) in \citet{Pfister2014} has been replaced with a computer. Instead of a human follower pressing a button in response to the participant's action, the computer speaks a voice command, using a voice which is clearly synthetic. We did not consider it realistic to represent a computer pressing a button without the use of robotics. A video of a button press might have been used, but we believe a synthetic voice command more clearly represented an action by the computer. Irrespective of the initial key pressed by the participant the computer responded by saying up, down, left, or right. In the general condition, the computer randomly selected the direction of the voice command. In the specific condition, the voice command corresponded to the arrow key pressed by the participant.

The overall sequence of each trial was as follows. First, an exclamation mark was shown on screen for 500 ms to indicate a trial is about to begin. Next, details were given of the interval the participant would estimate at the end of the trial. Participants then took an action by pressing either the space bar or an arrow key, depending on the condition, at a time of their own choosing. This action caused a beep sound. Following the beep, the computer spoke a voice command, which caused a square to move in the direction indicated by the computer. Participants' sense of agency was measured through intentional binding, using the interval estimation method. As illustrated in Figure \ref{st1:fig:round} measurements were taken for three different intervals:
    
\begin{itemize}
    \item Interval 1 - the interval between the participant's initial key press and the first beep sound. This varied in length from 200-700 ms, with the length randomised across trials. Measurements at this interval indicated participants' sense of agency in regard to their initial action and the first beep.
    \item Interval 2 - the interval between the beep sound and the computer voice command. This varied in length from 300-700 ms, with the length randomised across trials. 100 ms was added to the lengths in this interval to simulate the reaction time needed by the computer to issue a voice command. This reaction time was based on those of human participants in \cite{Pfister2014}. Measurements at this interval indicated participants' sense of agency in regard to the the computer voice command, which has been triggered by the beep.
    \item Interval 3 - the interval between the computer voice command and square moving. This varied in length from 200-700 ms, with the length randomised across trials. Measurements at this interval indicated participants' sense of agency over the onscreen movement caused by the computer voice command. i.e. the final outcome. 
\end{itemize}

While the intervals actually ranged in length from 200 ms to 700 ms, participants were told the intervals ranged in length from 50 ms to 1000 ms in steps of 50 ms, to allow them to over- or underestimate the interval lengths. Participants completed the trials by giving their interval estimate using onscreen radiobuttons, with options from 50 ms to 1000 ms in 50 ms increments.

The experiment involved 20 trials per interval, leading to a total of 60 rounds per condition, and 120 rounds overall (60 for general and 60 for specific commands). The order of the conditions and interval estimates was counterbalanced across participants. To help participants familiarise themselves with estimating interval lengths the experiment started with 40 training rounds. Here the intervals were equally distributed across intervals 1-3 and across conditions. Participants were told in advance which interval they had to estimate, allowing them to focus on the right interval during each trial. After estimating the interval in training rounds the participants were shown the actual length of the interval.

\subsection{Apparatus}
The study was conducted online using Testable \cite{rezlescu2020more}. Participants in our experiment required an Internet connection but completed the experiment unsupervised and in their own time. The experiment was implemented as a web application using the jsPsych Javascript library \cite{de2015jspsych} and could be accessed from any browser. Participants were asked to use a PC or laptop with a display, a hardware keyboard, a microphone, and a speaker or wired headphones. They were asked not to use Bluetooth headphones or speakers due to potential audio delays caused by the wireless technology. Participants were also instructed to participate in a calm environment, to turn off distracting notifications, and to run the experiment in full-screen mode for a consistent experience. We recognise that it is not possible for us to verify that participants followed these instructions. However, our approach is consistent with the one previously validated by \citet{Galang2021}.

\subsection{Participants}
Data was collected from 36 (8F, 28M) participants with a mean age of 26.44 (SD = 8.35). 
The vast majority of submissions came from India (26), with the rest coming from GB (2), Zambia (2), Australia (1), Ghana (1), Greece (1), Kenya (1), Philippines (1), and Turkey (1). The most common browser used was Chrome (33), with the other 3 participants using Edge.
Participation requirements included normal or corrected to normal vision and hearing. Participants were recruited through Testable \cite{rezlescu2020more}.

\subsection{Results}
\subsubsection{Data Cleaning}
Participants were tested for bias towards short or long interval estimations, which would mean they were choosing estimates mostly from one side of the screen. 
One participant was removed due to having estimations that were 2 standard deviations below the mean, suggesting that most guesses provided were very low regardless of the true interval length.
Previous research \cite{Pfister2014} has shown that the length of the interval being estimated is a strong predictor of estimation error.
This effect was confirmed through a linear regression, which found a significant negative effect of interval length on error. 
We decided to run further analysis on the regression residuals instead of the raw error scores in accordance with previous research \cite{Pfister2014} even though raw error estimates led to similar patterns of results. 

Trials, where the residual error was over 2.5 standard deviations away from the mean, were also removed, reducing the dataset by 1.82\%. This outlier removal process matches that of previous research using interval estimation \cite{Pfister2014}. Finally, to prepare for the analysis, the data was converted into participant means for each condition. Shapiro-Wilk tests confirmed the data was normally distributed and Mauchly's Test of Sphericity was not violated.

\subsubsection{Analysis}

\begin{figure}
    \centering
    \includegraphics[width=\linewidth]{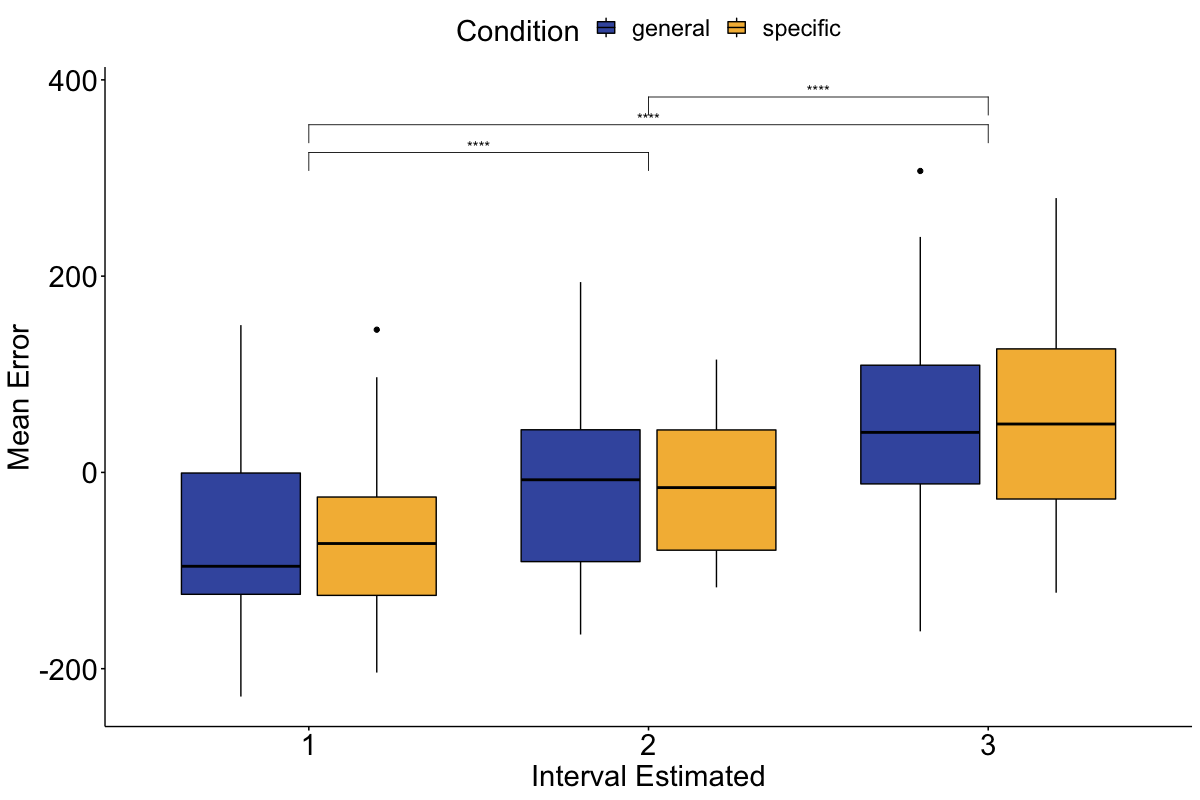}
    \caption{Boxplot of the mean error reported by participants for each interval and condition. Significance is shown as part of pairwise comparisons between intervals. Bonferroni correction applied. Negative scores in mean errors indicate the participants judged an interval as shorter than it actually was. Positive mean errors indicate the interval was judged as longer than it actually was. }
    \label{exp1.results}
\end{figure}

The data from 35 participants was included and analysed using R [version 4.1.3] and RStudio [version 2022.07.1+554 "Spotted Wakerobin" Release]. 
A two-way repeated measures analysis of variance (ANOVA) was conducted with \textit{interval} and \textit{condition} as independent variables, and \textit{mean error} as the dependent variable. 
The analysis showed a significant difference between \textit{intervals} [F(2, 68) = 28.28, p < 001], but not for \textit{condition} [F(1, 34) = .03, p = .87] or their interaction [F(2, 68) = .57, p = .57].
A subsequent post-hoc pairwise comparison with Bonferroni correction showed a significant increase in error for \textit{interval 3} when compared to \textit{interval 2} [t(69) = 5.13, p < .001] and \textit{interval 1} [t(69) = 8.88, p < .001]. There was also a significant increase in error between \textit{interval 1} and \textit{2} [t(69 = -4.86, p < 001].

To test whether intentional binding occurred for each interval, one sample tests were conducted comparing against a mean of zero. 
Interval 1 data was not normally distributed, failing a Shapiro-Wilk test, leading to a One-Sample Wilcoxon Signed Rank test. The result was significant, showing a marked decrease from 0 (V = 96.00, p < .001), which means participants experienced the interval as shorter than it actually was, thus indicating intentional binding occurred. As interval 2 was normally distributed, a one-sample t-test was conducted, finding no significant difference from 0 [t(34) = -.98, p = .333].
Lastly, the mean residual error in \textit{interval 3} was found to be significantly higher than zero [t(34) = 2.97, p = .005], meaning participants experienced this interval as longer than the actual interval.

\subsection{Discussion of Study 1} \label{st1:discussion}
Hypothesis 2 was not accepted as no significant differences were observed between the general and specific conditions. However, significant differences were observed in interval estimation errors across the three intervals. As described in Section \ref{relWork:SoA}, negative interval estimation errors - reflected by lower values in Figure \ref{exp1.results} - indicate a greater sense of agency. 

The estimation errors at interval 1 are negative and significantly different to zero. We can therefore conclude that the participants experienced a sense of agency over the initial beep caused by their key press. This result is consistent with prior studies, which have repeatedly shown that people experience a sense of agency in situations where a simple voluntary action (the key press) causes an outcome (the first beep). The estimation error at interval 3 was positive and significantly different to zero. We can thus conclude that participants did not experience agency in regard to the final movement of the square, which followed the computer's voice command. 
Based on these results, we can say with confidence that participants experienced a sense of agency over the initial beep caused by their key press, but did not experience agency over the final outcome of the task. This result is consistent with the results previously reported by \citet{Pfister2014} for human-to-human interaction.  

The estimation error at interval 2 is more difficult to interpret, as it is not significantly different to zero. However, significant differences were found between intervals 1 and 2, and also between intervals 2 and 3, with the error moving from negative (interval 1), to zero (interval 2), to positive (interval 3). 
This difficulty in interpreting interval 2 also complicates the interpretation of Hypothesis 1. 
Overall this indicates that participants experienced progressively less agency over events and outcomes as the task progressed.

The procedure in this study closely followed the protocol established by \citet{Pfister2014}. As noted above we believe that from an experimental design perspective this was the correct first step, providing a strong foundation for subsequent studies. However, from an HCI perspective, we also believe the procedure had two important limitations. Firstly the computer command was not a direct response to the participant's action. Instead, there was an intermediary event, a beep. The participant's keypress caused a beep, which in turn caused the voice command. In more typical interaction scenarios an intermediary event is unlikely. Secondly, the final movement of the square in Study 1 was not particularly meaningful. It made no real difference if the square moved up, down, left or right. In real-life situations, actions are completed with a purpose and usually result in different outcomes based on the action chosen. In Study 2 we address these limitations.

\section{Study 2} \label{sec:study2}
\begin{figure}
    \centering
    \includegraphics[width=\linewidth]{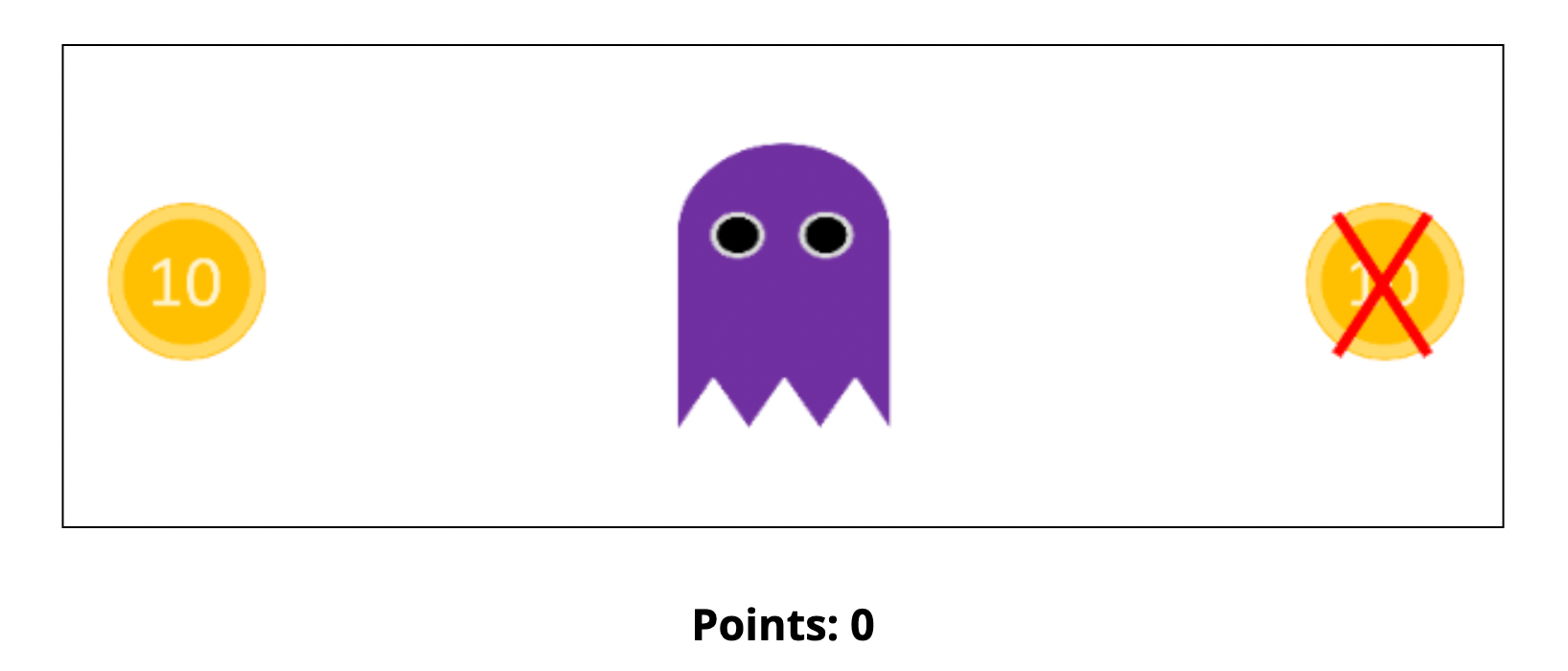}
    \caption{In study 2 a simple game was created to greater meaning to the experimental task. Participants gave instruction to move a character, that moved left or right to collect (symbol on the left) or lose (symbol on the right) 10 points.}
    \label{st2:fig:objectArea}
\end{figure}

Study 2 allowed us to address some of the limitations identified in Study 1. The first significant change was to introduce a more meaningful experimental task. This was done using a game-based scenario (Figure \ref{st2:fig:objectArea}). Instead of a non-descript square, the moving object is now a game character (shaped like an abstract ghost). The participant can collect or lose points based on the character's movements. Participants are told that their overall goal is to collect as many points as possible. Their accumulated points total is displayed onscreen. Whereas in Study 1 participants' choice of direction and the movement of the square was essentially random, our aim here was to provide an incentive by making the task more game-like.

We also simplified the experimental protocol in several ways. In Study 1 the square could move up, down, left, or right. Here the character moved only left or right. The second important change was the removal of the intermediate tone. In Study 1 the participant's action triggered a beep and the computer responded to this beep. In this experiment, the computer directly responded to the participant’s action. This means the participant’s command and the overall outcome (movement) were closer together in time. It also provides greater clarity in the overall sequence of actions and outcomes.

We again used Intentional Binding, measured through interval estimation, as an implicit measure for the sense of agency. But in a final significant change, we also asked participants to make an explicit judgment on agency over the final movement of the game character. This was done to investigate potential differences in participants' pre-reflective feeling and post-reflective judgement of agency.

We again had specificity as a condition in the study. In \textbf{\emph{the specific condition}} participants gave a left or right instruction using the left and right arrow keys and the computer repeated this instruction using a left or right voice command. In \textbf{\emph{the general condition}} participants pressed the spacebar, which generated a random left or right voice command. In both cases, the ghost character moved in the direction indicated by the computer voice command.  

This led to four conditions in total: general instruction with a positive outcome (\textit{generalPos}); general instruction with a negative outcome (\textit{generalNeg}); specific instruction with a positive outcome (\textit{specificPos}); specific instruction with a negative outcome (\textit{specificNeg}). As participants were aiming to accumulate as many points as possible, and \textit{specificNeg} caused them to lose points, we assumed they would not deliberately choose this option. When participants choose this option frequently, we interpreted it as meaning they either had not understood the instructions or were not paying attention to the task.

Study 2 had the following hypotheses:
\begin{enumerate}
    \item Participants will experience a sense of agency over the computer voice command in all conditions.
    \item Participants will experience a greater sense of agency over the final movement of the game character when their initial key press action is more specific.
\end{enumerate}

\subsection{Procedure}
\begin{figure}
    \centering
    \includegraphics[width=\linewidth]{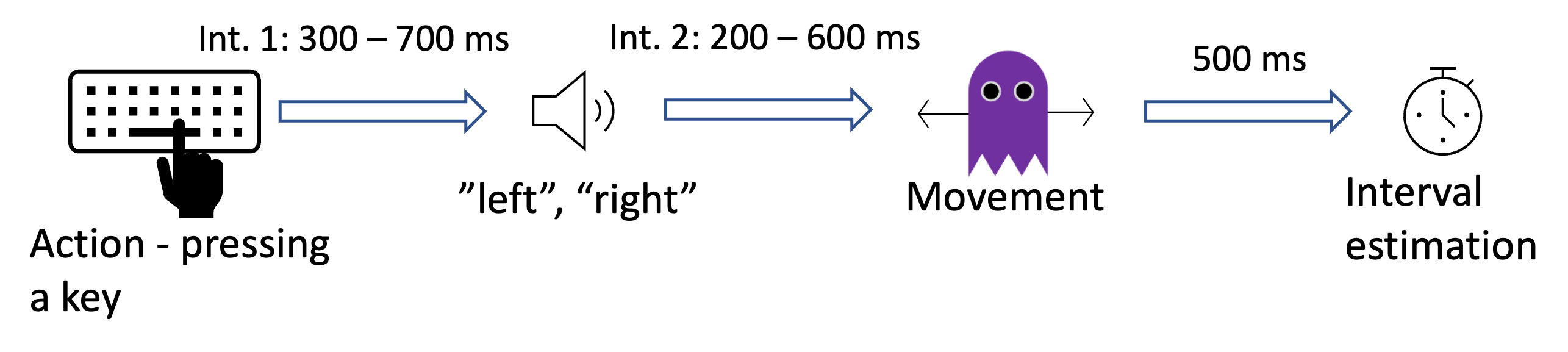}
    \caption{Procedure of each round - The participant starts by pressing a key on the keyboard, after 300 - 700 ms, the computer reacts to this by giving a voice command indicating a direction (left or right). After another 200 - 600 ms, the character, that is seen on screen, moves in the indicated direction. 500 ms after the movement is completed, the participants see a screen, where they are asked to estimate the length of one of the two intervals.}
    \label{st2:fig:round}
\end{figure}

Figure \ref{st2:fig:round} shows the procedure for each trial in this experiment. With the intermediary tone removed each trial had two intervals instead of three.
The computer reacted directly to the participant's action by giving the command "left" or "right". As before, the participants were able to choose a direction during the \textit{specific} condition, by pressing the left or right arrow key. In doing so, they were expected to move towards the reward, as their overall goal was to collect as many points as possible.
During the \textit{general} condition, the participant was not able to choose the direction of the movement and thus either gained or lost points randomly.
Following each trial, the participants were asked to estimate one of the two intervals. And after 18 trials, they were asked to judge their subjective experience of agency on a 5-point Likert scale.

\subsection{Apparatus}
This experiment used the same software as the first study and was also conducted online using Testable.

\subsection{Participants}
Data was collected from 50 participants (24F, 26M) recruited from the Testable online platform. Requirements for participation included: being 18 years of age or above and having normal or corrected vision and hearing. The mean age of participants was 31.69 (SD = 10.02), and they completed the study from a variety of different geographical locations. While most participants were from Great Britain (10), India (12), and Zambia (7), we also had submissions from Australia, Brazil, Ghana, Italy, Kenya, Portugal, Poland, the US, Venezuela, and others. Most participants used Chrome (45) to participate, with a smaller number using Edge (2) and Firefox (3).

As in Study 1 participants were asked to conduct the experiment in a calm environment, and use a hardware keyboard and wired headphones or speakers. Again we cannot verify if participants followed these instructions. 

\subsection{Results}
\subsubsection{Data Cleaning}
Data cleaning measures were similar to those of Study 1. One participant chose the negative option in the \textit{specific} condition more than 12 times, suggesting that they were not attempting to maximise rewards (as was requested in the instructions). 
As a result, this participant's data was removed. 
In addition, as the number of entries for \textit{specific negative} was lower than 5 per participant for the remaining participants, these entries were removed prior to analysis. 
In terms of outliers, two participants were removed due to being biased towards either very low or very high interval estimates, falling over 2 SD above those of other participants.
Again interval length was found to have a significant effect on estimation error, and so further analysis was conducted on the regression residuals.
Similarly to Study 1, we removed specific entries where the residual error in estimation was more than 2.5 z-scores away from the mean, which resulted in the removal of only 1.6\% of entries.

\begin{figure}
    \centering
    \includegraphics[width=\linewidth]{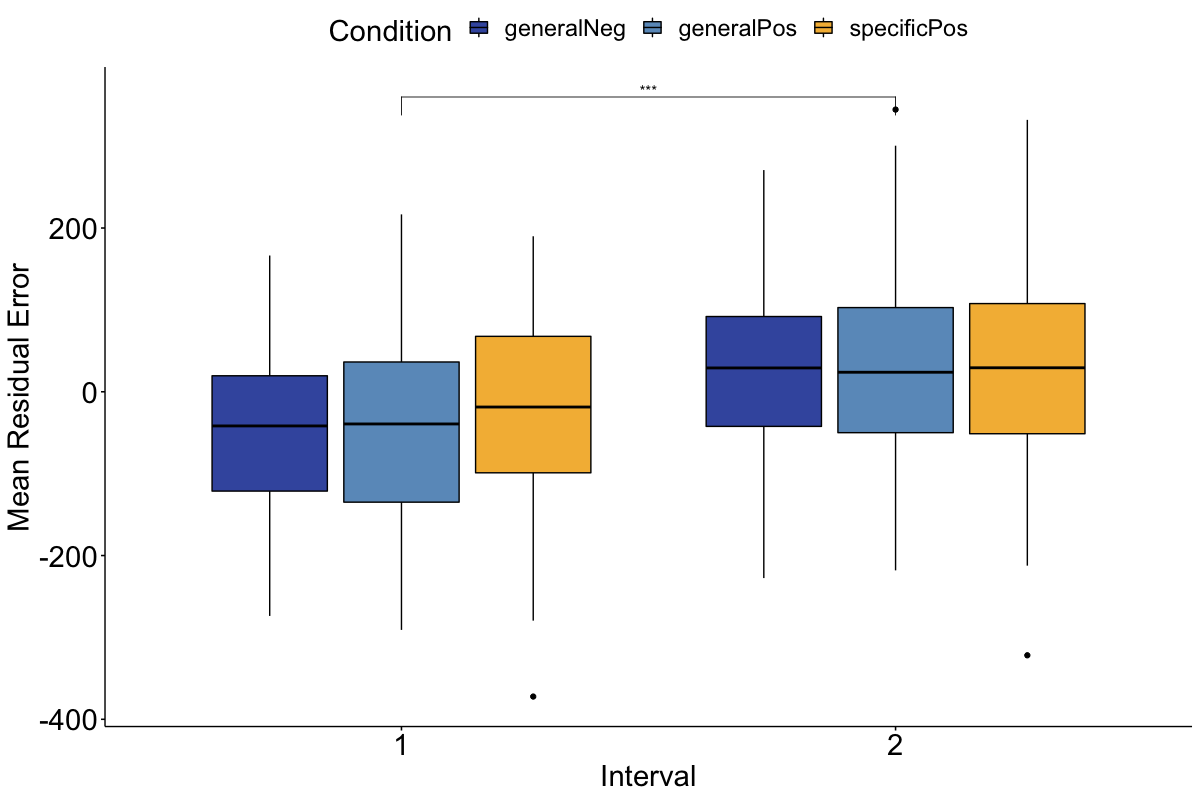}
    \caption{Boxplot of the mean error reported by participants for each interval and condition. Significance is shown as part of the t-test between intervals.}
    \label{exp2.results}
\end{figure}

\subsubsection{Analysis}
After testing for relevant assumptions, a two-way repeated measures ANOVA was run on the means of each participant per \textit{condition} and per \textit{interval}. The analysis showed a significant effect of \textit{interval} [F(1, 46) = 74.55, p < .001], but not of \textit{condition} [F(2, 92) = 0.71, p = .32; GG corrected], or of their interaction [F(2, 92) = .80, p = .44; GG corrected]. A subsequent t-test of \textit{interval} showed that mean residual error for \textit{interval 2} (M = 31.7, SD = 126) was significantly higher than for \textit{interval 1} (M = -41.9, SD = 115); [t(140) = -6.81, p < .001]. Additionally, a one-sample t-test was used to test for intentional binding for each interval. Error residuals for \textit{interval 1} were found to be significantly lower than zero (M = -41.90, SD = 100.52); [t(46) = -2.86, p = .006], suggesting intentional binding occurred. No significant difference was found for \textit{interval 2} (M = 30.62, SD = 108.52); [t(46) = 1.93, p = .059], indicating no binding. Figure \ref{exp2.results} provides an overview of these results.

\begin{figure}
    \centering
    \includegraphics[width=\linewidth]{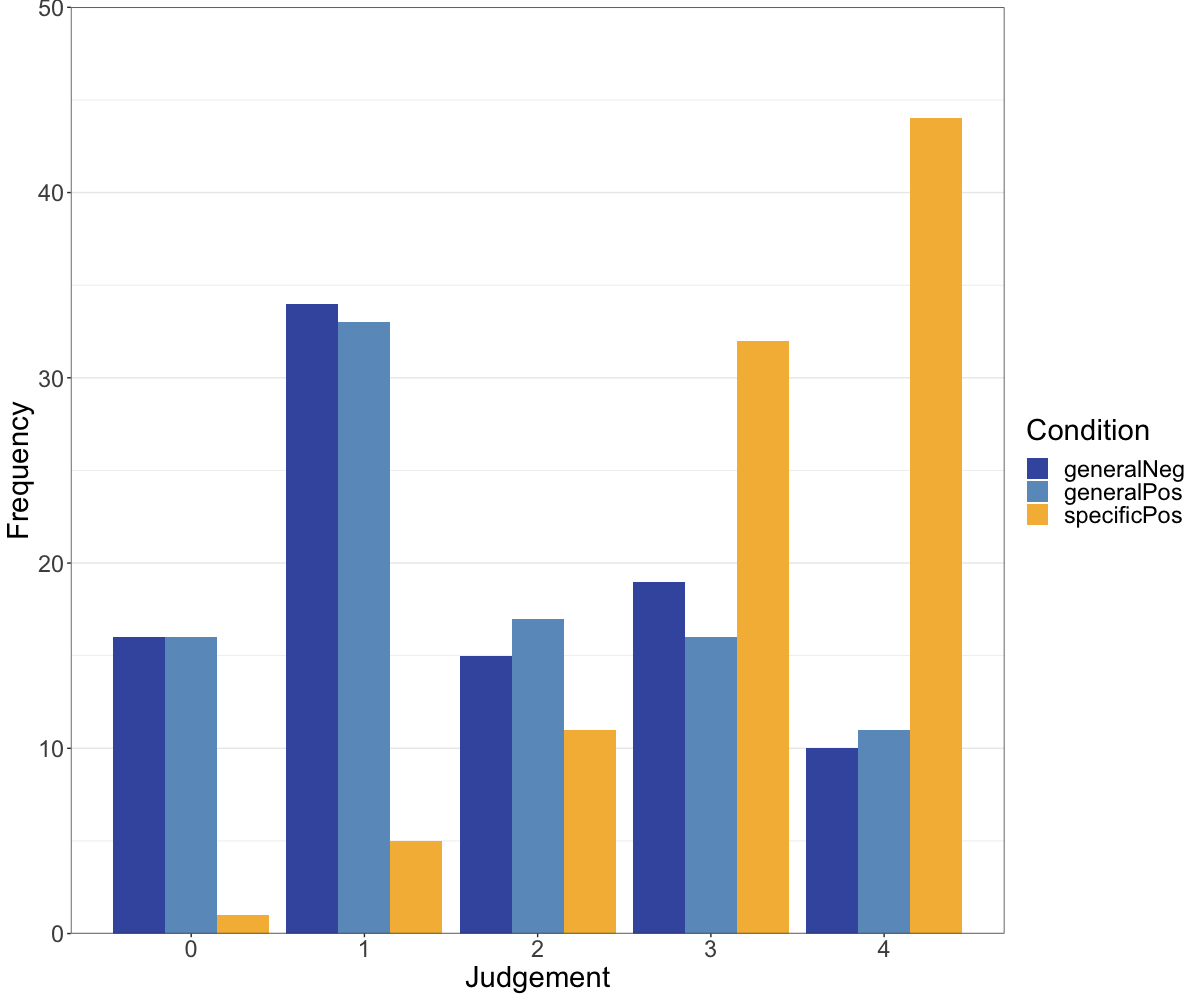}
    \caption{Barplot showing the frequency of judgements per condition. The Chi-square test showed a significant difference due to condition.}
    \label{judge.results}
\end{figure}

Participants were asked to judge how much they believed themselves to be in control of the movement ("To which extent do you feel like you caused the character to move in the last round?") seen on screen from 0 - 4, with 4 being fully in control ("It was me who caused the movement") and 0 being not in control ("I did not cause the movement"). See Figure \ref{judge.results} for results. A chi-square analysis showed a significant difference due to condition on judgement. Examining the chi-square residuals showed that participants rated themselves to be mostly in control during the \textit{specific positive} condition, while mostly not in control for the two \textit{general} conditions.

\subsection{Discussion of Study 2}
Study 2 addressed limitations of the first study, introducing a more meaningful task, and making the computer’s voice command an immediate reaction to the participant’s action. Regarding the more meaningful task, results show that when participants could specify the direction of movement for the character, they almost always chose to collect points (in 99.23 \% of the trials in the specific condition), rarely choosing the \textit{specificNeg} option. This indicates that their choice of a left or right key press was deliberate and that they engaged with the overall aim of collecting as many points as possible. We thus conclude that participants saw their actions as meaningful (at least within the context of the experimental task, but please refer to Section \ref{sec:limits} for further discussion).

For interval 1 participants' estimation errors were negative and significantly different to zero. We thus conclude that intentional binding occurred and that participants experienced a pre-reflective feeling of agency over the computer voice command triggered by their keypress. This means we can accept hypothesis 1.
In contrast, no intentional binding effect was observed at interval 2 (estimation errors of zero). This indicates participants did not experience pre-reflective agency over the character's movement. Overall we can conclude that while participants had a pre-reflective feeling of agency over the computer's voice command, they did not have this feeling for the character's movement. The significant difference between estimation errors at intervals 1 and 2 reinforces this conclusion.

No significant differences were observed in interval estimation errors between the general and specific condition.
We thus conclude that the specificity of participants' actions had no impact on their pre-reflective feeling of agency.
However, a different picture emerges when we examine the results of participants' post-reflective judgement of agency.
Here we find that the specificity of the keyboard command did have an impact. Hence, we can only partly accept Hypothesis 2 \footnote{With the benefit of experimental hindsight we also now realise that Hypothesis 2 might have been better stated as two separate hypotheses, one focusing on the feeling of agency, and the second addressing judgement.}.

When asked to explicitly report their judgement of agency, participants generally reported strong judgements of having caused the character's movement in the \textit{specificPos} condition. This suggests participants did indeed judge themselves to be in control in situations where they had a choice over the direction the character would move. 
In both general conditions, participants reported weak judgements of control. This indicates that the ability to choose a specific action, rather than a positive or negative outcome, led to the judgement of being in control. 
Overall, we can conclude that participants judged themselves to 
be agentic when they could choose a specific and meaningful action.

\section{General Discussion} \label{sec:discussion}
This paper has explored people's sense of agency in a situation where they issued an  instruction to a computer. We have explored the sense of agency using both implicit and explicit measures, and thus provide insight into both pre-reflective feelings of agency and post-reflective judgements of agency. We will now consider both the theoretical and design implications of our findings.

\subsection{Theoretical Implications}
Both studies in the paper suggest that people did not experience a pre-reflective feeling of agency over final outcomes that resulted from their instructions to a computer. This is consistent with the findings of similar studies on human-human interaction \cite{Pfister2014, Caspar2018}. From a theoretical perspective, it is interesting to consider this finding in light of \citet{reeves1996media}'s research on the \textit{Media Equation}, which argues that people attribute human characteristics to computers and media. They suggest that people tend to behave as if they are interacting with another human, even when actually interacting with a computer \cite{reeves1996media}, an idea captured in the Computer as Social Actors (CASA) framework \cite{nass2000machines}.

Other researchers have also found instances where people treat computers in a manner similar to humans, particularly when interacting through speech. For example, a study looking at accented synthetic speech on lexical choice found that similar to how humans interact with each other, participants will change the words they use to align with the accent of synthetic speech produced by a computer \cite{Cowan2019}. People also adopt politeness strategies when interacting with computers, including asking for permission to speak or apologising to the computer agent for an error \cite{large2017steering}. Even without the use of speech interfaces, participants will be kinder when reviewing a computer if they are entering their answers on the same computer, compared to a different computer or paper and pen questionnaire \cite{nass1999people}, again providing evidence that people transfer certain qualities of human-to-human interaction to their interactions with computers. However, this transfer is not complete, and CASA theory does not suggest that people treat computers exactly like humans on every occasion. In fact, studies usually show differences in degree, but can also produce differences in responses altogether \cite{edwards2016robots, sundar2019machine}. Our studies explored the sense of agency in HCI from a command-agent perspective. In the case of pre-reflective feelings of agency, our results are consistent with those predicted by the Media Equation and provide further evidence for this theory.

\subsection{Design Implications}
As described in the related work section, the feeling of agency is a pre-reflective or instinctive experience. While the feeling of agency is often unconscious it has a direct bearing on our day-to-day life, providing our background experience of control and causality, and providing a stable foundation for our actions in the world. In contrast, the judgement of agency is post-reflective. It happens after the event, when we deliberately evaluate our actions, or those of others, and attribute responsibility or causality for an outcome.

Participants in Study 2 did not experience a pre-reflective feeling of agency over the final movement of the on-screen character. But under certain conditions, they did judge themselves to have caused the movement (a post-reflective judgement of agency). This happened when they were involved in a meaningful task and could specify the direction the character should move to. While prior research has reported differences in implicit and explicit measures of agency \cite{Synofzik2008, Dewey2014, Obhi2011, stetson2006motor}, ours is the first to do so when examining agency from a command-agent perspective in an HCI context.

The results of Study 2 have important implications for design. The sense of agency is closely linked to our sense of causality and responsibility. The findings of Study 2 suggest that, where we want people to have a sense of control and take responsibility for outcomes that result from instructions they give to a computer, we cannot rely on pre-reflective or instinctive feelings of agency. Instead we should design systems that encourage people to reflect on their actions and make deliberative judgments about responsibility. Previous HCI research has argued for the benefits of slow technology \cite{hallnas2001slow}, design friction \cite{cox2016design}, error prevention \cite{gould2016input}, and uncomfortable interactions \cite{benford2012uncomfortable} in the context of cultural computing and in support a greater reflection and sense of wellness. Our results suggest that "friction" could also be a useful design tool in situations where we want people to reflect and actively consider their personal responsibility when giving instructions to computers.

\section{Limitations} \label{sec:limits}
Both studies in this paper were conducted online. As noted in Section 4 we are aware of concerns and challenges related to online experiments \cite{9463290}. This is something we considered carefully before undertaking our studies. Previous research examining the quality of experimental data collected online and in-person found no discernible difference \cite{casler2013separate, Galang2021}, leading the authors to recommend online research for behavioural and psychological measures. This included evidence that speaks directly to the validity of asynchronous online studies using intentional binding as a measure of the sense of agency \cite{Galang2021}. Together these studies give us confidence in the validity of our methods.

Our first study closely followed the protocol established in prior psychology research on the sense of agency by \citet{Pfister2014}. Its key function was to provide robust foundations for future studies. From an experimental design perspective, we believe this was a rigorous approach to take. However, as noted in Section \ref{st1:discussion}, we believe Study 1 had limitations from an HCI perspective, which were addressed in Study 2. The fact that we did not identify a pre-reflective feeling of agency over final outcomes in either experiment gives us confidence in our results.

In a relative sense, the task in Study 2 was more meaningful than in Study 1. Results show participants acted to maximise the number of points collected, suggesting a meaningful connection between their actions and their overall goal of collecting the largest number of points possible. However, we also accept that the experimental task involved simple actions and outcomes. We are therefore careful to couch our findings with this limitation in mind. Without further evidence, we cannot prove that these results will apply to more meaningful interactions such as autonomous command and control systems, interactions with personal agents, or even computer-supported remote surgery. Future studies are needed to explore scenarios in which people's instructions to computers come with greater consequences. We argue our work is an important first step in addressing such situations.

\section{Conclusion} \label{sec:conclusion}
This paper reported two studies, each focused on an individual's sense of agency from a social processing perspective in a situation where they give instructions to a computer which acted on their behalf.
Our results indicate that while participants experienced a pre-reflective feeling of agency over the action taken by the computer, they did not feel agentic over the final outcome, that movement. 
It was further found that the participant's command specificity and the outcome did not influence this instinctive sense of agency. These results replicated similar studies on human-human interaction \cite{Pfister2014}, extending the research to the HCI context and supporting the idea of computers as social actors \cite{nass2000machines}. 
In addition, we found that participants judged themselves to be the cause of the movement when able to control the outcome, but not when the outcome was chosen by the computer. 
This finding highlights the differences between pre-reflective feelings of agency and post-reflective judgements of agency, suggesting that these measures are less sensitive to different processes. 
This difference shows that we should design systems in a way, that enables people to reflect on their actions and to make a judgement about their responsibility.
Future work is needed to further examine the factors that affect each type of agency, so that we may better understand which measure is important in each new context. This knowledge would serve to improve interactions with computers, allowing interfaces to be built around the idea of fostering user agency and improving user sense of control.

\section{Acknowledgements}
This work was conducted with the financial support of the Science Foundation Ireland Centre for Research Training in Digitally-Enhanced Reality (d-real) under Grant No. 18/CRT/6224.
Support was also provided by Science Foundation Ireland through the Adapt Research Centre (13/RC/2106\_P2) and Insight Centre for Data Analytics (12/RC/2289\_P2).

\bibliographystyle{elsarticle-num-names} 
\bibliography{bibliography.bib}





\end{document}